\documentclass[twocolumn,showpacs,preprintnumbers,amsmath,amssymb]{revtex4}
\usepackage{graphicx}
\usepackage{dcolumn}
\usepackage{bm}
\usepackage{mathrsfs}

\newcommand{\rd}{{\rm d}}
\newcommand{\re}{{\rm e}}

\begin{document}

\preprint{APS/123-QED}

\title{Bound states of spatial optical dark/gray solitons in nonlocal media}

\author{Shigen Ouyang}
\author{Wei Hu}
\author{Qi Guo}
\email{guoq@scnu.edu.cn} \affiliation{Laboratory of Photonic
Information Technology, South China Normal University Guangzhou,
510631, P. R. China}

\date{\today}

\begin{abstract}
It is shown that three or more dark/gray solitons can form bound
states in nonlocal media. More over dark/gray solitons can form
bound states in several balance distances. Numerical simulations
indicate that some of such bound states are unstable and will
decay into a group of fundamental solitons, while others may be
stable. There exist degenerate bound states with the same
velocity, Hamiltonian, particle numbers and momentum but decaying
in different ways and having different lifetimes.
\end{abstract}


\maketitle

\section{Introduction}
As were pointed out previously by N. I. Nicolov, et. al.\cite{1}
and Y. V. Kartashov, et. al.\cite{2}, two dark/gray solitons can
form bound states due to their long-range attraction in nonlocal
nonlinear media. In this paper we will show that three or more
dark/gray solitons can form bound states also. As were indicated
previously\cite{3}, nonlocal dark/gray solitons have exponentially
decaying oscillatory tails which in turn give rise to widely
distributed exponentially decaying oscillatory light-induced
perturbed refractive index. As a result, in this paper, we will
find that dark/gray solitons can form bound states in several
balance distances. Numerical simulation shows that fundamental
dark/gray solitons are stable. Some of such bound states are
unstable and will decay into a group of fundamental solitons,
while others may be stable. Bound states of no central symmetry or
asymmetry will have degenerate states with the same non-vanishing
velocity, Hamiltonian, particle numbers and momentum. Some of such
degenerate states are unstable and will decay in different ways
and have different lifetimes.

\section{bound states of multiple dark/gray solitons}
The propagation of a paraxial optical beam in a medium with a
self-defocusing spatially nonlocal nonlinearity can be described
by the dimensionless nonlocal nonlinear Schr\"{o}dinger
equation(NNLSE)\cite{1,2,3}
\begin{equation}\label{1-1}
i{{\partial u}\over{\partial z}}+{{1}\over{2}}{{\partial^2
u}\over{\partial x^2}}-u\int R(|x-x^\prime|)|u(x^\prime,z)|^2\rd
x^\prime=0
\end{equation}
where $u(x,z)$ is the complex amplitude envelop of the light beam,
$I(x,z)=|u(x,z)|^2$ is the light intensity, $x$ and $z$ are
transverse and longitude coordinates respectively, $R(x)$ is the
real nonlocal response function and satisfies the normalization
condition $\int R(x)\rd x=1$, $n(x,z)=-\int
R(|x-x^\prime|)|u(x^\prime,z)|^2\rd x^\prime$ is the light-induced
perturbed refractive index. Note that not stated otherwise all
integrals in this paper will extend over the whole $x$ axis. It is
easy to prove if $u(x,z)$ satisfies Eq.~(\ref{1-1}) with a
nonlocal response function $R(x)$, the scale-transformed function
$\widetilde{u}(x,z)={{1}\over{\alpha}}u({{x}\over{\alpha}},{{z}\over{\alpha^2}})$
satisfies Eq.~(\ref{1-1}) with another nonlocal response function
$\widetilde{R}(x)={{1}\over{\alpha}}R({{x}\over{\alpha}})$, where
$\alpha$ is an arbitrary positive real number. So it is adequate
to consider dark/gray soliton having asymptotic behavior
$|u(x,z)|^2\xrightarrow{x\rightarrow\pm\infty}1$. Not stated
otherwise all dark/gray solitons in this paper have this
asymptotic behavior.

By introducing a transformation $u(x,z)=\psi(x,z)\re^{-iz}$,
equation~(\ref{1-1}) turns into
\begin{eqnarray}
i{{\partial\psi}\over{\partial z}}+{{1}\over{2}}{{\partial^2
\psi}\over{\partial x^2}}-\psi\left[\int
R(|x-x^\prime|)|\psi(x^\prime)|^2\rd
x^\prime-1\right]=0\label{1-2}
\end{eqnarray}
which has three integrals of motion\cite{4,5}, namely, the number
of particles
\begin{eqnarray}
N=\int(1-|\psi|^2)\rd x,
\end{eqnarray}
the momentum
\begin{eqnarray}
P={{i}\over{2}}\int\left(\psi{{\partial\psi^*}\over{\partial
x}}-\psi^*{{\partial\psi}\over{\partial
x}}\right)\left(1-{{1}\over{|\psi|^2}}\right)\rd x,
\end{eqnarray}
and the Hamiltonian
\begin{eqnarray}
&&H={{1}\over{2}}\int\Big|{{\partial\psi}\over{\partial
x}}\Big|^2\rd x+\int(1-|\psi|^2)\rd x~~~~~~~~~~~~~~~~~~\nonumber\\
&&~~~~+{{1}\over{2}}\int\!\!\!\int
R(|x-x^\prime|)\Big[|\psi(x)|^2|\psi(x^\prime)|^2-1\Big]\rd x\rd
x^\prime~~~
\end{eqnarray}

In this paper we numerically solve Eq.~(\ref{1-2}) to find
dark/gray soliton solutions
\begin{eqnarray}
\psi(x,z)=\phi(x-v z)\re^{i\theta(x-v z)},\label{soliton}
\end{eqnarray}
where $v$ is the velocity of the dark/gray soliton relative to the
cw background, and real functions $\phi$ and $\theta$
asymptotically approach
$\phi(x)\xrightarrow{x\rightarrow\pm\infty}1$,
$\theta(x)\xrightarrow{x\rightarrow\pm\infty}\mp\theta_0$, where
$\theta_0$ is a real constant. By inserting Eq.~(\ref{soliton})
into (\ref{1-2}), we get
\begin{subequations}\label{base}
\begin{eqnarray}
&\theta^\prime=v\left(1-{{1}\over{\phi^2}}\right)\\
&{{\phi^{\prime\prime}}\over{2}}+{{v^2+2}\over{2}}\phi-{{v^2}\over{2\phi^3}}-\phi\int
R(x-x^\prime)\phi^2(x^\prime)d x^\prime=0
\end{eqnarray}
\end{subequations}

As an example, we consider this following nonlocal case in which
the light-induced perturbed refractive index is governed by
\begin{eqnarray}
n-w^2{{\partial^2 n}\over{\partial x^2}}=-|u|^2,\label{n}
\end{eqnarray}
which results in $ n(x,z)=-\int
R(|x-x^\prime|)|u(x^\prime,z)|^2\rd x^\prime$, where
$R(|x|)={{1}\over{2w}}\exp\left(-{{|x|}\over{w}}\right)$ and $w$
is the characteristic nonlocal response length of the media.
\begin{figure}
\centering
\includegraphics[width=1in]{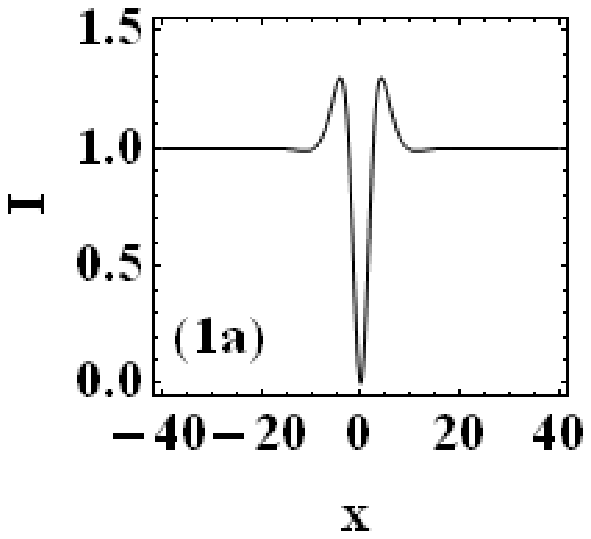}
\includegraphics[width=2in]{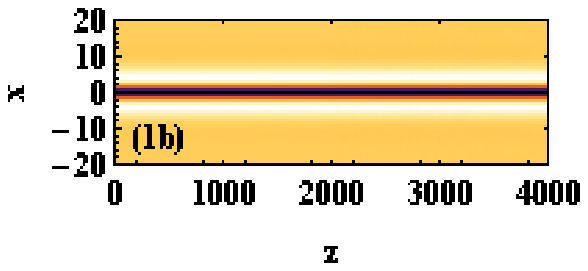}\\
\includegraphics[width=1in]{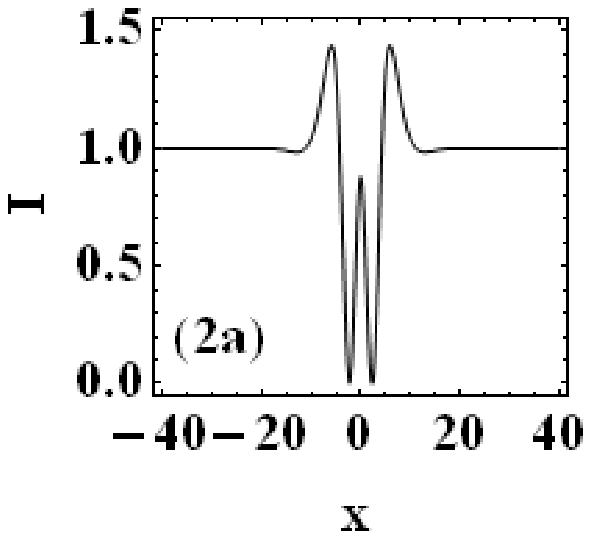}
\includegraphics[width=2in]{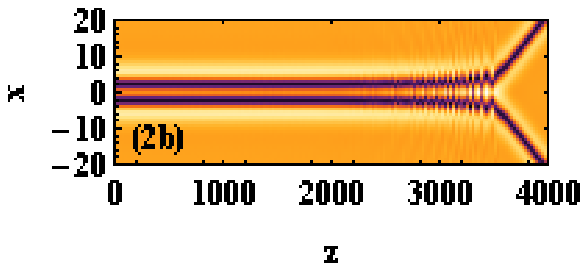}\\
\includegraphics[width=1in]{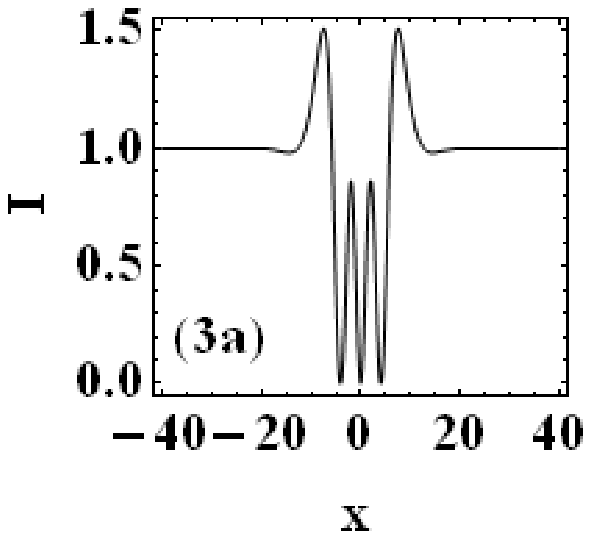}
\includegraphics[width=2in]{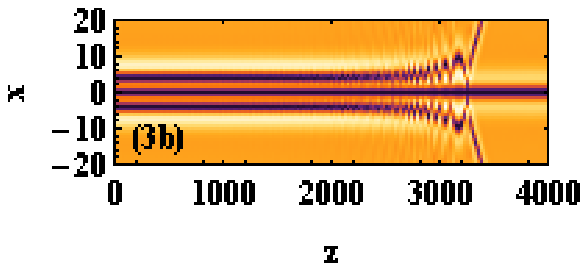}\\
\includegraphics[width=1in]{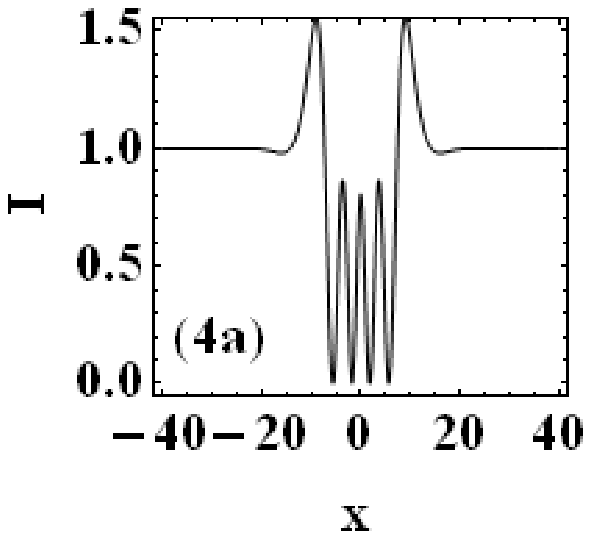}
\includegraphics[width=2in]{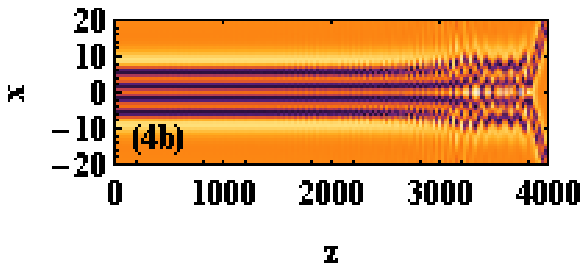}
\caption{\label{I}(1a),(2a),(3a),(4a) are the intensity profiles
of $\psi_0$,$\psi_1$,$\psi_{1,1}$,$\psi_{1,1,1}$ respectively.
(1b),(2b),(3b),(4b) are their counterpart evolutions. Here $w=5,
v=0$.}
\end{figure}
Numerically solving Eqs.~(\ref{base}), we can find the dark and
gray soliton solutions. The intensity profiles of dark solitons
$\psi_0$ and $\psi_1,\psi_{1,1},\psi_{1,1,1}$ and their
counterpart evolutions are shown in Fig.~(\ref{I}) when $w=5,
v=0$. (Similarly, bound states $\psi_1,\psi_{1,1},\psi_{1,1,1}$
can form on the interface of two meida\cite{9}). No observable
changes in the intensity profile of the fundamental soliton
$\psi_0$ can be found during its propagation . Numerical
simulation (not shown here) indicate that an initially broadened
beam $\psi(x,0)=\psi_0({{x}\over{1.5}},0)$ will converge into
$\psi_0$ quickly during its propagation. So the fundamental
soliton $\psi_0$ is stable. However, from Fig.~(\ref{I}), with no
initial perturbation bound states $\psi_1,\psi_{1,1},\psi_{1,1,1}$
are all unstable and will decay into a group of fundamental
solitons. Numerical simulations (not shown here) also indicate
that initially broadened bound states, instead of converging into
bound solitary states, will ultimately decay into a group of
fundamental solitons.

\begin{figure}
\centering
\includegraphics[width=1in]{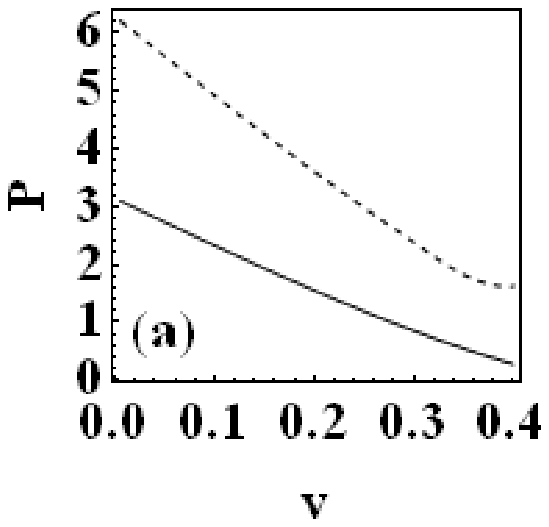}
\includegraphics[width=1.1in]{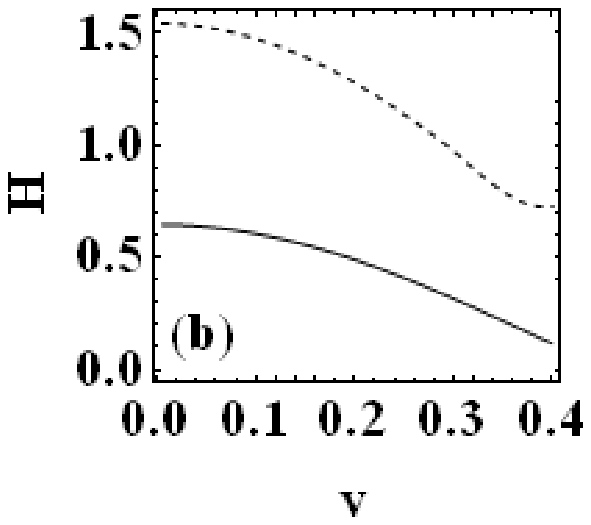}
\includegraphics[width=1.1in]{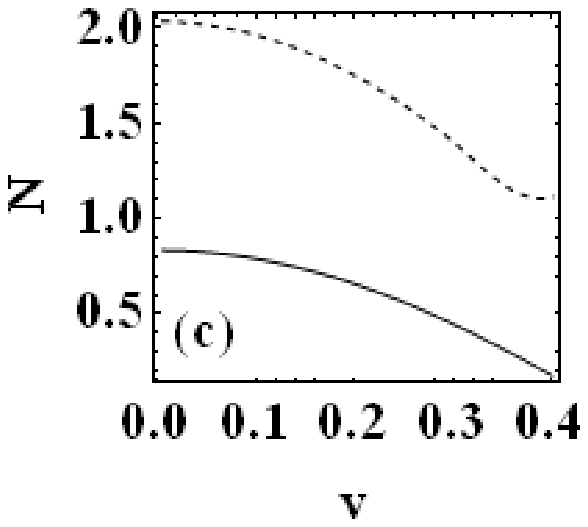}
\caption{\label{II}Functional dependence of momentum $P$,
Hamiltonian $H$ and the number of particles $N$ on the velocity
$v$ of $\psi_0$ (solid line) and $\psi_1$ (dashing line) when
$w=5$.}
\end{figure}

As has been shown previously\cite{3},when $w>1/2$ the maximal
velocity of gray soliton is $v_{\rm
max}=\sqrt{{4w-1}\over{4w^2}}$. So when $w=5$, we have $v_{\rm
max}=0.436$. As shown in Fig.~(\ref{IV})(1b), when $w=5$ the bound
state $\psi_1$ with an initial velocity $v=0.4$ is unstable.
Numerical simulations (not shown here) also indicate that $\psi_1$
with other velocities $v=0.1,0.2,0.3$ are all unstable. So we
strongly guest bound state $\psi_1$ with any velocity are unstable
when $w=5$. Similarly, bound state $\psi_{1,1},\psi_{1,1,1}$ with
any velocity are all unstable when $w=5$. When $w=5$, the
functional dependence of momentum $P$, Hamiltonian $H$ and
particle numbers $N$ on the velocity $v$ of $\psi_0$ and $\psi_1$
are shown in Fig.~(\ref{II}), from which, both for $\psi_0$ and
$\psi_1$, we get ${\rm d }P/{\rm d}v<0$ for all velocity. $\psi_0$
is stable but $\psi_1$ is not. So the criterion of dark soliton
instability ${\rm d }P/{\rm d}v>0$\cite{6,7,8} may be a sufficient
but not necessary condition. Namely we cannot tell a dark/gray
soliton whether stable or not if ${\rm d }P/{\rm d}v<0$.

\begin{figure}
\centering
\includegraphics[width=0.7in]{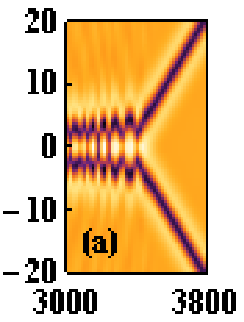}
\includegraphics[width=0.7in]{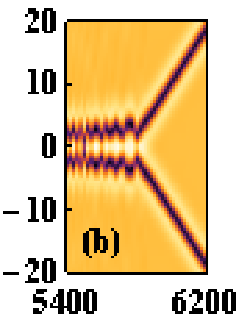}
\includegraphics[width=0.7in]{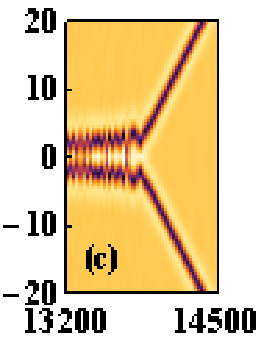}
\includegraphics[width=0.7in]{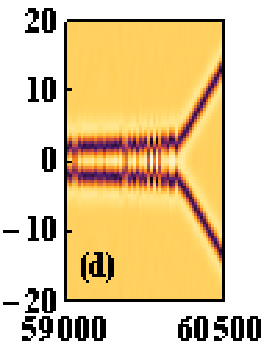}
\caption{\label{III}It needs a longer and longer propagation
distance for the decaying of bound states $\psi_1$ as the
characteristic nonlocal response length decreases from (a) $w=4$
to (b) $w=2$, to (c) $w=1.5$, and to (d) $w=1.2$.}
\end{figure}

As is shown in Fig.~(\ref{III}), it needs a longer and longer
propagation distance for bound states $\psi_1$ with $v=0$ to decay
into fundamental solitons as the characteristic nonlocal response
length $w$ decreases from $w=4$ to $w=2$, and to $w=1.5$, and to
$w=1.2$. So it is possible that bound states $\psi_1$ could be
stable for a small enough value of $w$. But limited by the nature
of the numerical simulations method used by this paper we cannot
present an exact proof for the stability of such bound state. So
it is still an open question of the stability of bound states of
dark/gray solitons for small value of $w$.

\section{Balance distances between dark/gray soltions}
\begin{figure}
\centering
\includegraphics[width=1in]{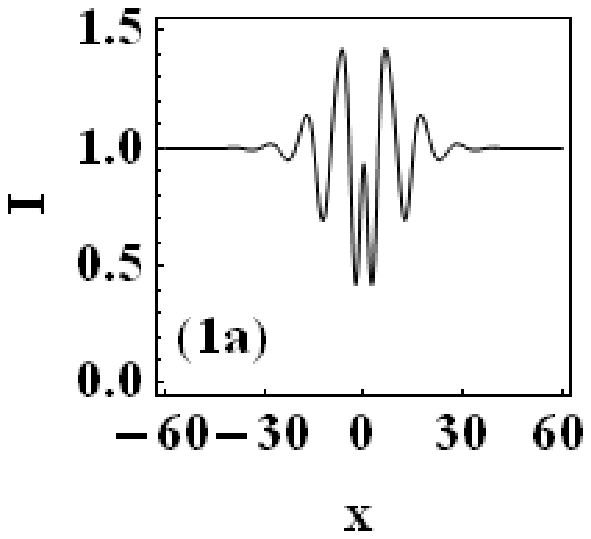}
\includegraphics[width=2in]{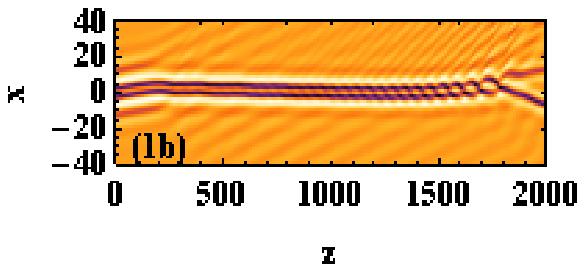}\\
\includegraphics[width=1in]{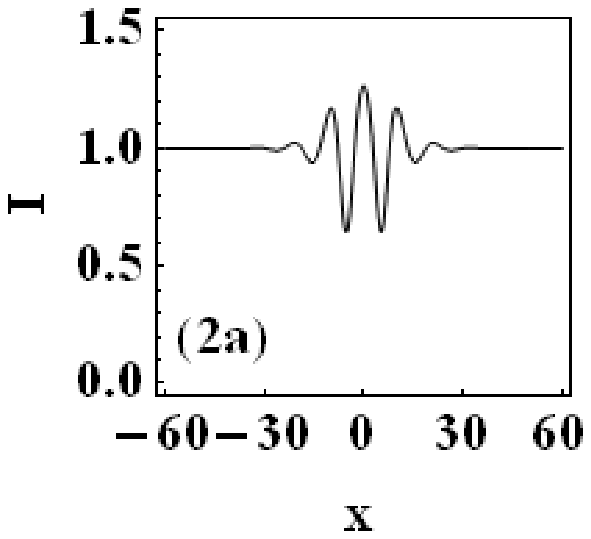}
\includegraphics[width=2in]{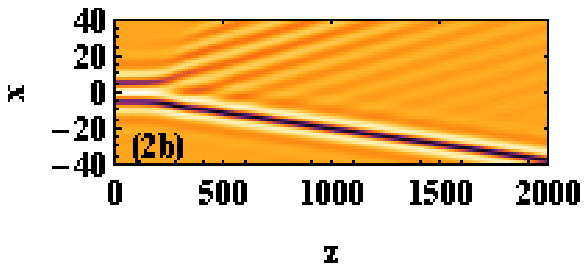}\\
\includegraphics[width=1in]{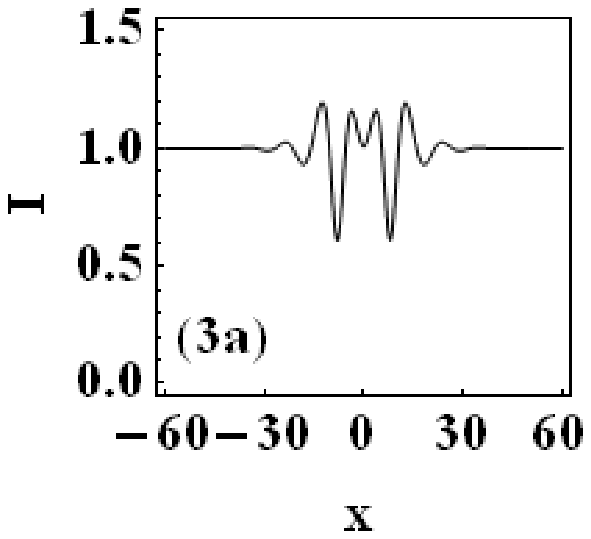}
\includegraphics[width=2in]{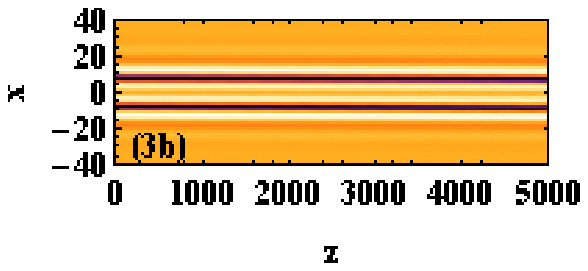}\\
\includegraphics[width=1in]{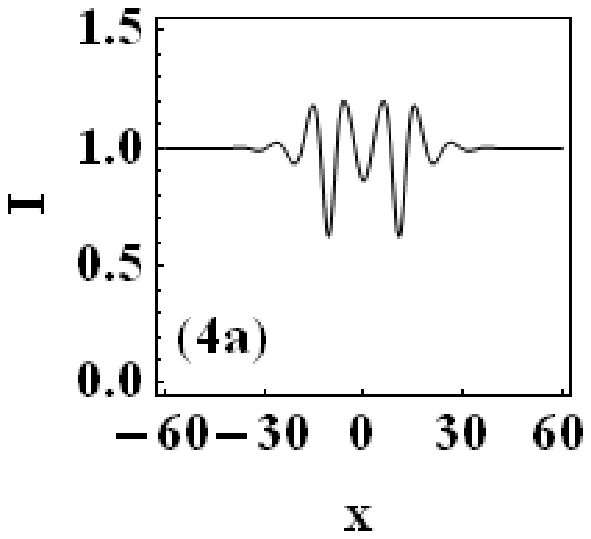}
\includegraphics[width=2in]{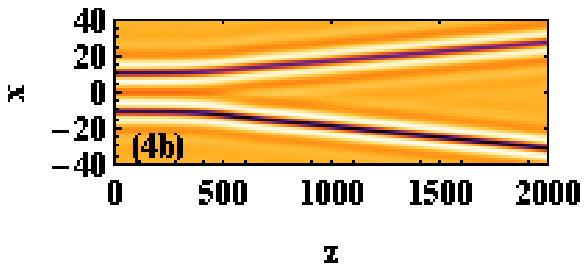}\\
\includegraphics[width=1in]{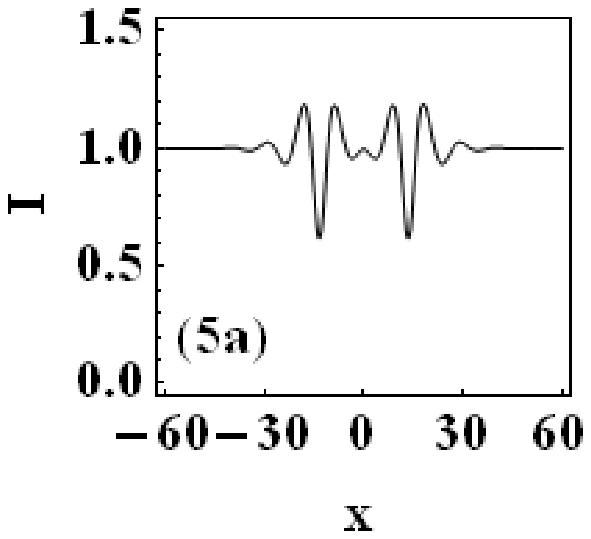}
\includegraphics[width=2in]{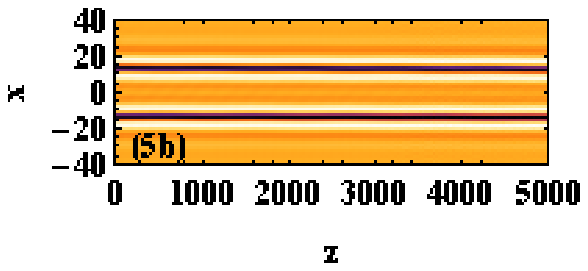}\\
\includegraphics[width=1in]{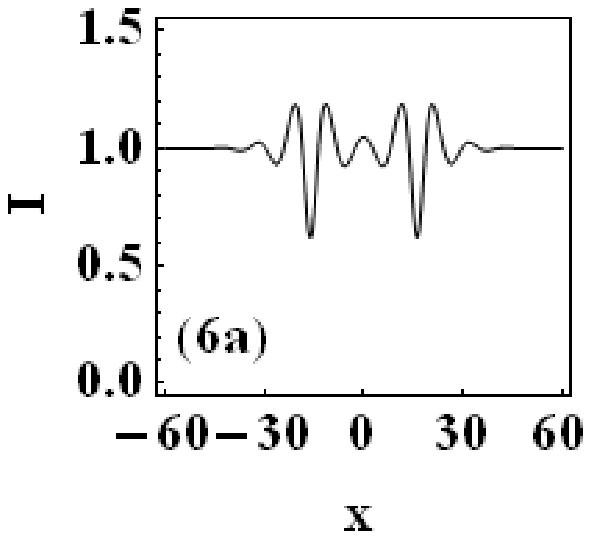}
\includegraphics[width=2in]{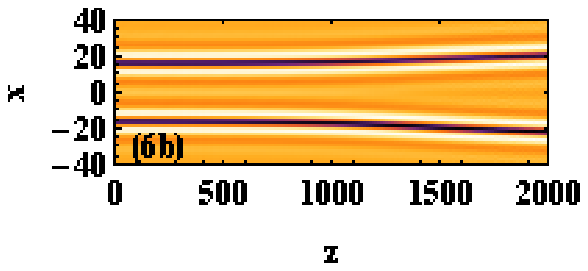}\\
\includegraphics[width=1in]{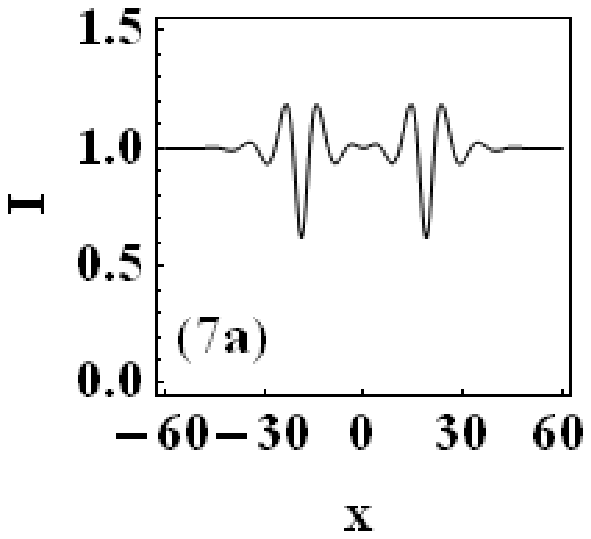}
\includegraphics[width=2in]{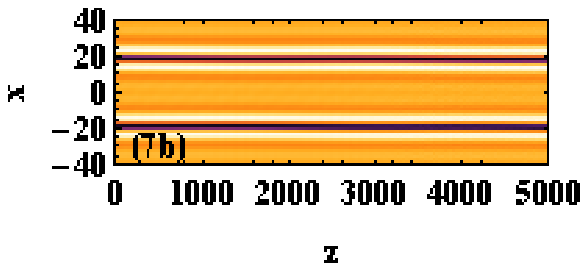}\\
\includegraphics[width=1in]{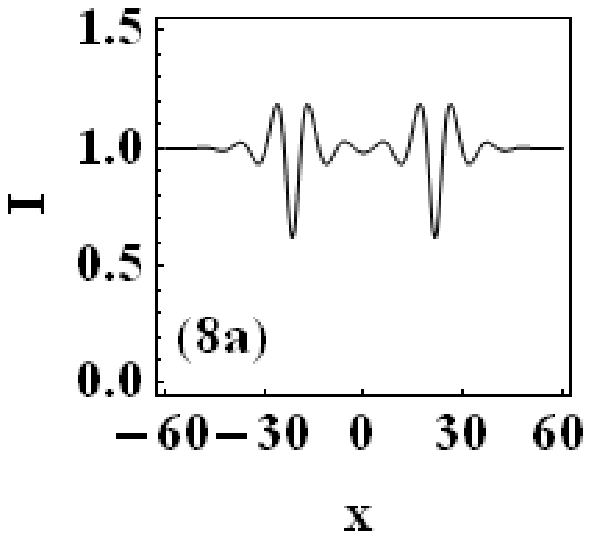}
\includegraphics[width=2in]{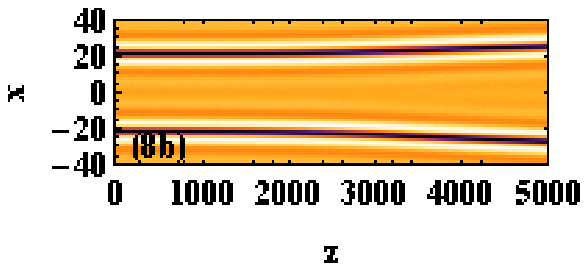}\\
\caption{\label{IV}(1a),(2a),(3a),(4a),(5a),(6a),(7a),(8a) are the
intensity profiles of
$\psi_1$,$\psi_2$,$\psi_3$,$\psi_4$,$\psi_5$,$\psi_6$,$\psi_7$,$\psi_8$
respectively. (2b),(3b),(4b),(5b),(6b),(7b),(8b) are their
counterpart evolutions figured in frames moving with velocity
$v=0.4$ and (1b) figured in a frame moving with $v=0.38$. Here
$w=5$ and velocities of bound states are all $v=0.4$.}
\end{figure}
\begin{figure}
\centering
\includegraphics[width=1.4in]{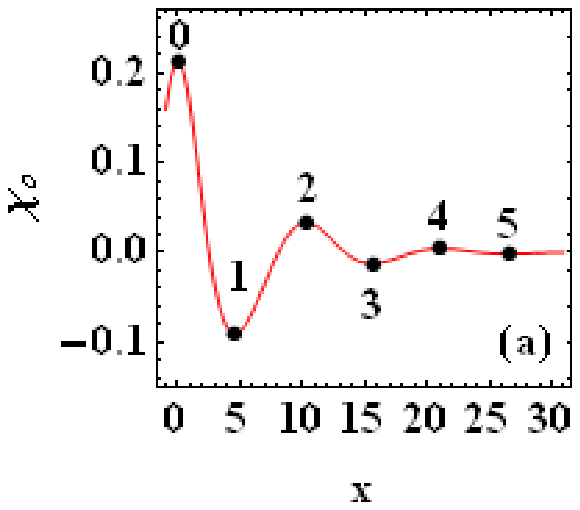}
\includegraphics[width=1.47in]{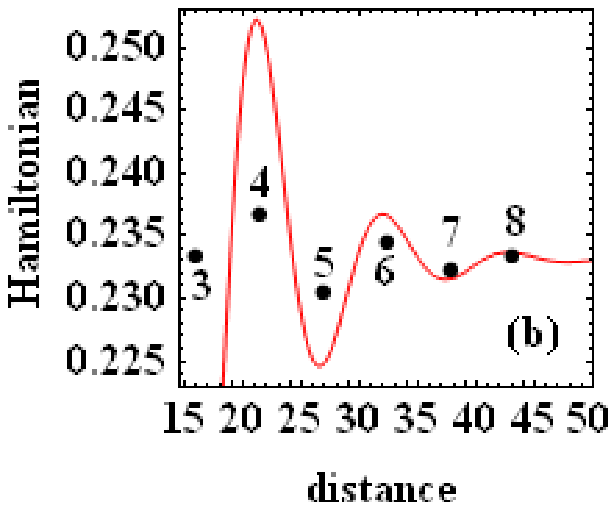}
\includegraphics[width=1.4in]{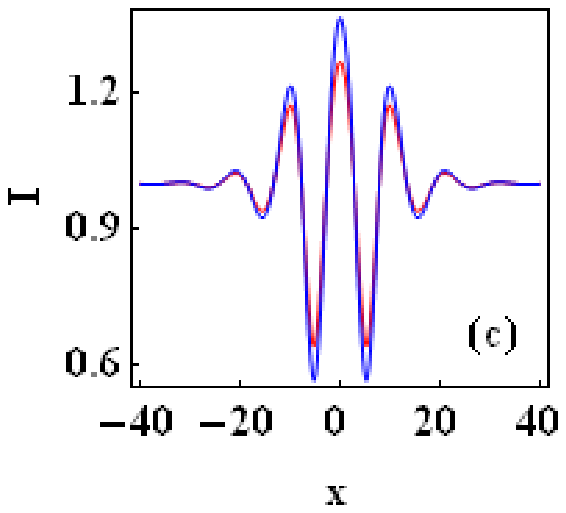}
\includegraphics[width=1.4in]{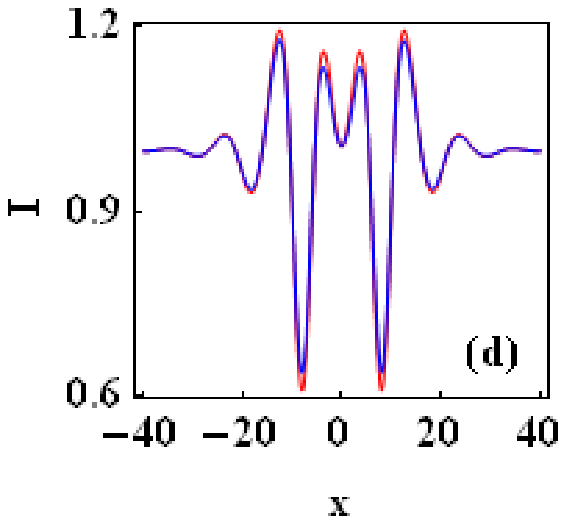}
\caption{\label{V}(a)fundamental soliton $\chi_0(x)$ (red line)
and its maximums and minimums (black dots). (b) the Hamiltonian of
$\psi_d$ (red line) and those of
$\psi_3,\psi_4,\psi_5,\psi_6,\psi_7,\psi_8$ (black dots). (c),(d)
the intensity profiles of $\psi_2,\psi_3$ (red line) and
$\psi_{d_2},\psi_{d_3}$ (blue line). Here $w=5,v=0.4$.}
\end{figure}

As was indicated previously\cite{3}, due to the nonlocal nonlinear
response of the media the dark/gray solitons have exponentially
decaying oscillatory tails when $|v|\leq\sqrt{{4w-1}\over{4w^2}}$
for $w>{{1}\over{4}}$. Let $\phi(x)=1-\chi(x)$, then when
$|x|\rightarrow\infty$, we have\cite{3}
\begin{eqnarray}
\chi(x)\approx c_1\exp(-\lambda_1|x|)\cos(\lambda_2|x|+c_2)
\end{eqnarray}
where $c_1$ and $c_2$ are two constants, and
\begin{eqnarray}
&&\lambda_1=\sqrt{{1-4w^2v^2+4w\sqrt{1-v^2}}\over{4w^2}},\\
&&\lambda_2=\sqrt{{-1+4w^2v^2+4w\sqrt{1-v^2}}\over{4w^2}},\label{lambda2}
\end{eqnarray}
and $2\pi/\lambda_2$ is the oscillatory spatial period. As shown
in Fig.~(\ref{V})(a), when $w=5, v=0.4$, the fundamental soliton
$\chi_0$ has a serial of maximums and minimums interdigitally
located at
$x_0=0,x_1=4.57,x_2=10.29,x_3=15.61,x_4=21.10,x_5=26.52,x_6=31.97,x_7=37.41,x_8=42.85$.
Such exponentially decaying oscillatory tails in turn give rise to
an exponentially decaying oscillatory light-induced perturbed
refractive index, and dark/gray solitons can form bound states in
not only one but several balance distances. For example, as shown
in Fig.~(\ref{IV}), when $w=5,v=0.4$, the balance distances (the
distance between two deepest dips) of bound states
$\psi_1$,$\psi_2$,$\psi_3$,$\psi_4$,$\psi_5$,$\psi_6$,$\psi_7$,$\psi_8$
are
$d_{i}=4.88$,$10.48$,$15.97$,$21.33$,$26.82$,$32.24$,$37.69$,$43.13$,
where index $i$ runs from 1 to 8. Obviously we have $d_i\approx
x_i$. Let $\Delta d_i\equiv d_{i+1}-d_i$, we have $\Delta
d_i=5.6$,$5.49$,$5.36$,$5.49$,$5.42$,$5.45$,$5.44$. On the other
hand, from Eq.~(\ref{lambda2}), the half of the oscillatory
spatial period $\pi/\lambda_2=5.44$ very closes to $\Delta d_i$.
Other cases with different $w$ and $v$ also show the same relation
between $\pi/\lambda_2$ and $\Delta d_i$.
To study the interaction of two fundamental solitons $\psi_0$
separated by a distance $d$, we introduce a coupled state
$\psi_d(x)=\phi_d(x)\exp[i\theta_d(x)]$, where
\begin{subequations}
\begin{eqnarray}
&\phi_d(x)=1-\chi_0\left(x-{{d}\over{2}}\right)-\chi_0\left(x+{{d}\over{2}}\right),\\
&\theta_d^\prime=v\left(1-{{1}\over{\phi_d^2}}\right),
\end{eqnarray}
\end{subequations}
and the index $d$ of $\psi_d$ denotes the distance. Obviously,
when $d\rightarrow\infty$, there will be no overlap between
$\chi_0(x-{{d}\over{2}})$ and $\chi_0(x+{{d}\over{2}})$, and
$\psi_d$ will decouple into two well separated fundamental
solitons $\psi_0$. We note, when $d=x_1, x_3, x_5, \cdots$, the
maximums of $\chi_0(x-{{d}\over{2}})$ will overlap with the
minimums of $\chi_0(x+{{d}\over{2}})$, and as Fig.~(\ref{V})(b)
shows, the Hamiltonian of $\psi_{d}$ takes the minimums, while
when $d=x_2, x_4, x_6, \cdots$, the maximums of
$\chi_0(x-{{d}\over{2}})$ will overlap with the maximums of
$\chi_0(x+{{d}\over{2}})$, and the Hamiltonian of $\psi_{d}$ takes
the maximums. On the other hand, as indicated by
Fig.~(\ref{V})(c),(d), the difference between $\psi_3$ and
$\psi_{d_3}$ is small, and the difference between $\psi_i$ and
$\psi_{d_i}$ will decrease as $d_i$ increases. So, as
Fig.~(\ref{V})(b) shows, though there is a difference between the
Hamiltonian of $\psi_{d_i}$ and those of $\psi_i$, the Hamiltonian
of $\psi_i$ assumes a similar behavior of that of $\psi_d$, which
may qualitatively explain why $d_i\approx x_i$ and $\Delta
d_i\approx\pi/\lambda_2$. Numerical simulations shown in
Fig.~(\ref{IV}) indicate that $\psi_2,\psi_4,\psi_6,\psi_8$ are
all instable, while $\psi_3,\psi_5,\psi_7$ may be all stable, and
$\psi_1$ is weakly instable when $w=5,v=0.4$.

\begin{figure}
\centering
\includegraphics[width=0.9in]{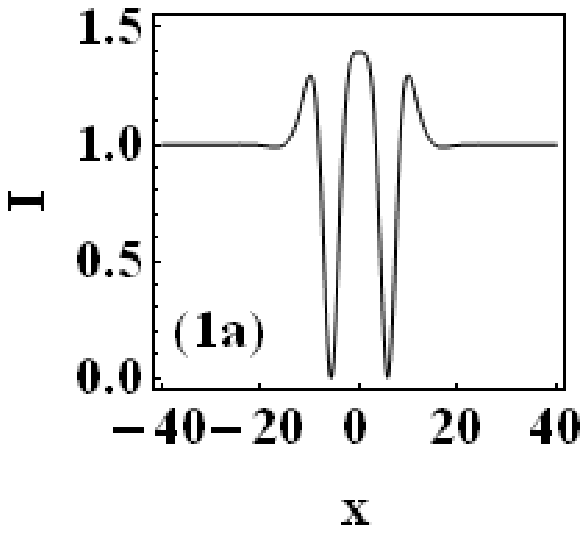}
\includegraphics[width=1.2in]{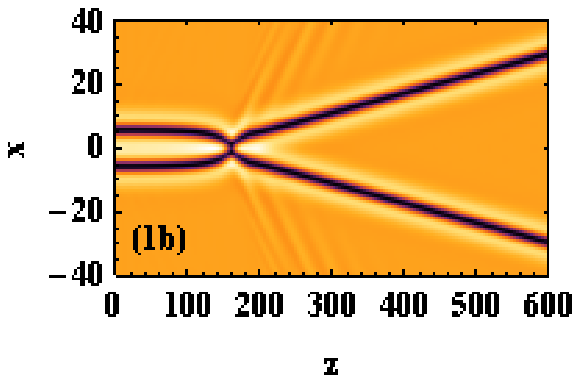}
\includegraphics[width=1.2in]{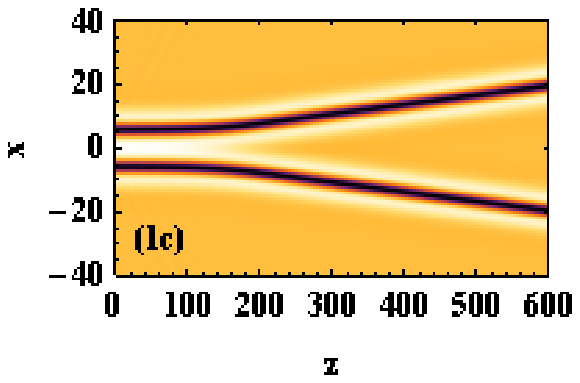}\\
\includegraphics[width=0.9in]{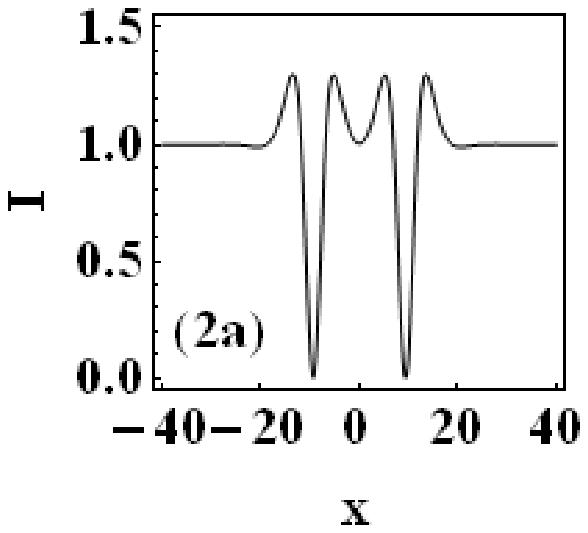}
\includegraphics[width=1.2in]{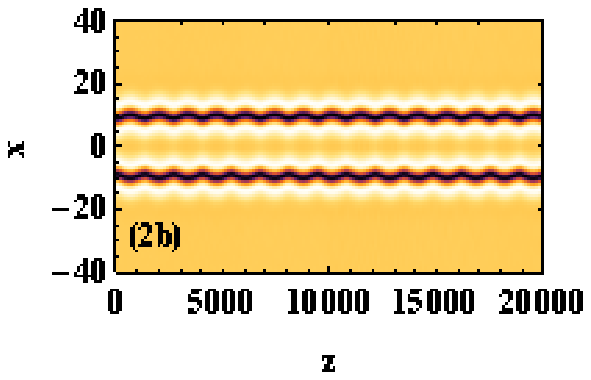}
\includegraphics[width=1.2in]{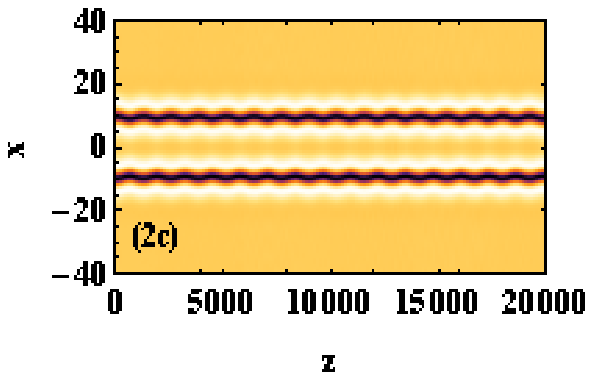}\\
\includegraphics[width=0.9in]{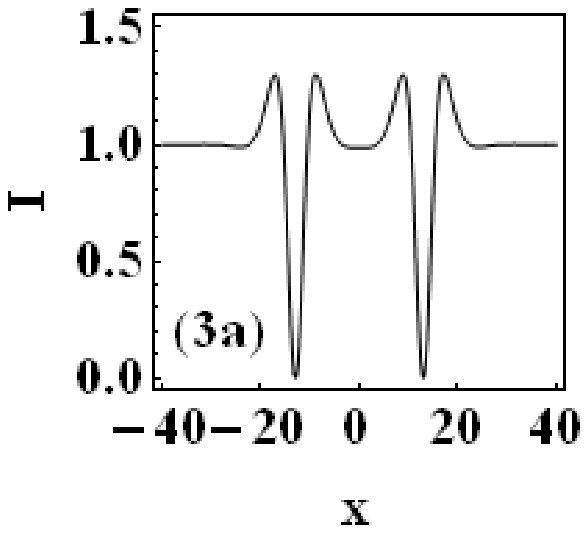}
\includegraphics[width=1.2in]{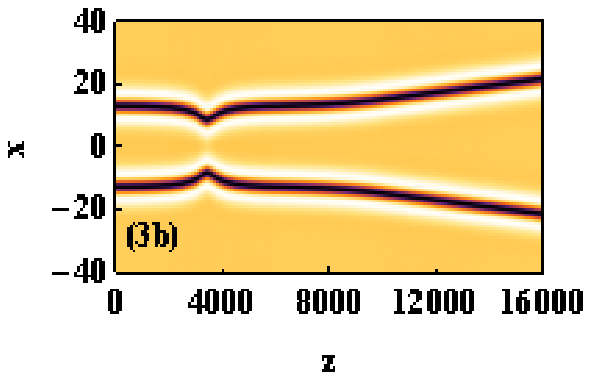}
\includegraphics[width=1.2in]{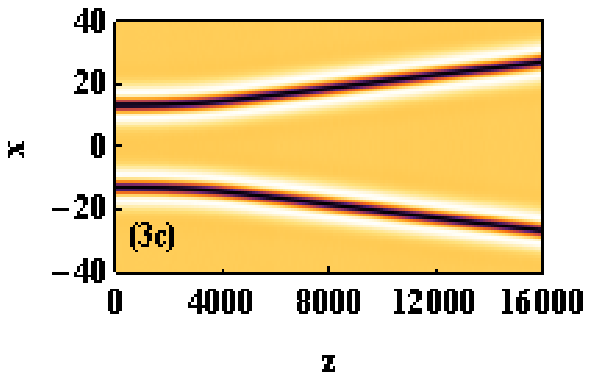}
\caption{\label{VI}(1a),(2a),(3a) are intensity profiles of
$\psi_2,\psi_3,\psi_4$; Numerical simulations of initially
broadened bound states
(1b)$\psi_2({{x}\over{0.99}})$,(1c)$\psi_2({{x}\over{1.01}})$;
(2b)$\psi_3({{x}\over{0.95}})$,(2c)$\psi_3({{x}\over{1.05}})$;
(3b)$\psi_4({{x}\over{0.99}})$,(3c)$\psi_4({{x}\over{1.01}})$
respectively. Here $w=5,v=0$}
\end{figure}

Numerical simulations with initially broadened bound states with
vanishing velocity $v=0$, seeing Fig.~(\ref{VI}), also indicate
that slightly initial departure from the bound states $\psi_2$ and
$\psi_4$ can result in a great different evolutions in the long
run. $\psi_2$ and $\psi_4$ are both instable when $w=5,v=0$ also.
On the other hand the initially broadened bound state of $\psi_3$,
instead of converging into $\psi_3$ or decaying into two
fundamental solitons, will fall in a seemingly eternal (over a
distance larger than $z=100000$) vibrations around $\psi_3$. So
bound state $\psi_3$ may be oscillatory-stable. But it needs a
more rigorous proof beyond the numerical simulation method of this
paper to judge the stability of $\psi_3,\psi_5,\psi_7$.

\begin{figure}
\centering
\includegraphics[width=0.9in]{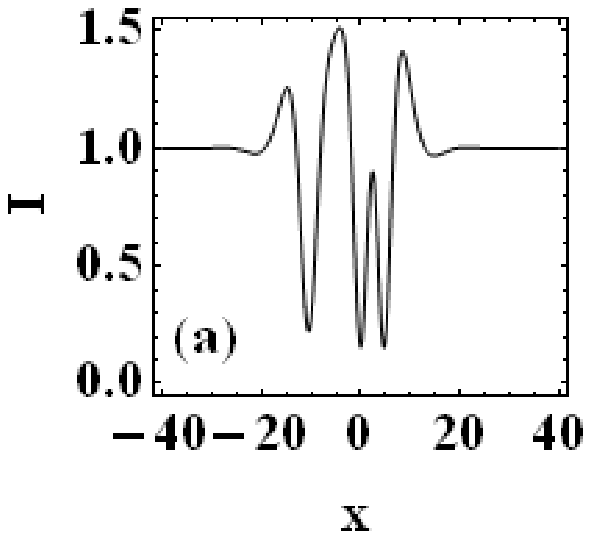}
\includegraphics[width=1.2in]{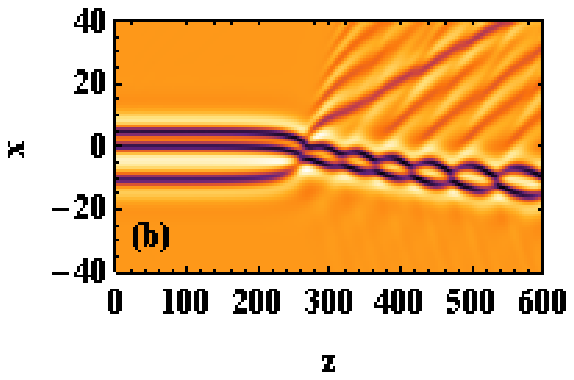}
\includegraphics[width=1.2in]{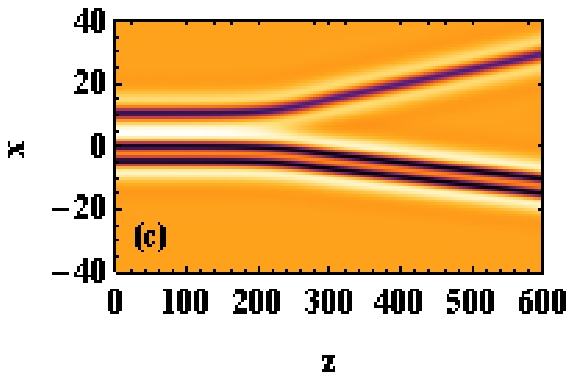}
\caption{\label{VII}(a) The intensity profile of $\psi_{2,1}$; Two
degenerate bound states (b) $\psi_{2,1}$ and (c) $\psi_{1,2}$
decay in different ways. Here $w=5,v=0.25$.}
\end{figure}

Since dark/gray solitons can form bound states at several balance
distances, we can construct bound states like $\psi_{1,2,1}$ or
$\psi_{\psi_{3,3}}$, and so on. Interestingly, as shown in
Fig.~(\ref{VII}), bound states with no central symmetry or
asymmetry will have two degenerate states, like $\psi_{1,2}$ and
$\psi_{2,1}$, moving with the same non-vanishing velocity and
having the same momentum, Hamiltonian and the particle numbers but
decaying in different ways and having different life-times.
However, there may still exist stable degenerate states, like
$\psi_{3,5}$ and $\psi_{5,3}$.

\section{conclusion}
In nonlocal media, multiple dark/gray solitons can form bound
states in several balance distances. Some of such bound states are
unstable and will decay into a group of fundamental solitons,
while others may be stable. There exist degenerate bound states
with the same velocity, Hamiltonian, particle numbers and momentum
but decaying in different ways and having different lifetimes.

\begin{acknowledgments}
This research was supported by the National Natural Science
Foundation of China (Grant No. 61008007) and Specialized Research
Fund for growing seedlings of the Higher Education in Guangdong
(Grant No. LYM10066).
\end{acknowledgments}

　　　　　　　　　　　　　　
\end{document}